\documentclass[useAMS,usenatbib]{mnras}

\usepackage{url} 
\usepackage{subfig}
\usepackage{graphicx}
\usepackage{amsmath}

\newcommand\ltsim{\mathrel{\substack{
			\textstyle<\\[-0.2ex]\textstyle\sim}}}

\usepackage{multirow}
\graphicspath{{fig/}} 

\usepackage[utf8]{inputenc}
\usepackage[english]{babel}

\def\aj{AJ}
\def\apj{ApJ}
\def\apjl{ApJ}
\def\apjs{ApJS}
\def\ao{Appl.~Opt.}
\def\aap{A\&A}
\def\mnras{MNRAS}
\def\pasa{PASA}
\def\pasp{PASP}

\bibpunct{(}{)}{,}{a}{}{,} 

\title[Opacity, variability and kinematics of AGN jets]{Opacity, variability and kinematics of AGN jets}

\author[Kutkin et al.]{\parbox{\textwidth}{
A.~M.~Kutkin$^{1,2}$\thanks{E-mail: kutkin$\star$asc.rssi.ru},
I.~N.~Pashchenko$^{1}$, 
K.~V.~Sokolovsky$^{1,3}$, 
Y.~Y.~Kovalev$^{1,4,5}$,
M.~F.~Aller$^{6}$,
H.~D.~Aller$^{6}$
}
\vspace{0.4cm}\\
\parbox{\textwidth}{
$^{1}$Astro Space Center of Lebedev Physical Institute, Profsoyuznaya Str.
84/32, 117997 Moscow, Russia\\
$^{2}$ASTRON, Netherlands Institute for Radio Astronomy, Oude Hoogeveensedijk 4, 7991PD, Dwingeloo, The Netherlands\\
$^{3}$Sternberg Astronomical Institute, Moscow State University,
Universitetskii~pr. 13, 119992 Moscow, Russia\\
$^{4}$Moscow Institute of Physics and Technology, Dolgoprudny, Institutsky per., 9, Moscow region, 141700, Russia\\
$^{5}$Max-Planck-Institut f\"ur RadioAstronomie, Auf dem H\"ugel 69, 53121 Bonn, Germany\\
$^{6}$University of Michigan, 311 West Hall, Dept of Astronomy, Ann Arbor MI, 48109-1107, USA\\
}
}

\begin{document}

\date{Accepted XXX. Received YYY; in original form ZZZ}

\pagerange{\pageref{firstpage}--\pageref{lastpage}} \pubyear{2019}

\maketitle

\label{firstpage}

\begin{abstract}
 Synchrotron self-absorption in active galactic nuclei (AGN) jets manifests itself as a time delay between flares observed at high and low radio frequencies. It is also responsible for the observing frequency dependent change in size and position of the apparent base of the jet, aka the core shift effect, detected with very long baseline interferometry (VLBI).
 We measure the time delays and the core shifts in 11 radio-loud AGN to estimate the speed of their jets without relying on multi-epoch VLBI kinematics analysis.
 The 15--8\,GHz total flux density time lags are obtained using Gaussian process regression, the core shift values are measured using VLBI observations and adopted from the literature.
 A strong correlation is found between the apparent core shift and the observed time delay.
 Our estimate of the jet speed is higher than the apparent speed of the fastest VLBI components by the median coefficient of 1.4. The coefficient ranges for individual sources from 0.5 to 20. We derive Doppler factors, Lorentz factors and viewing angles of the jets, as well as the corresponding de-projected distance from the jet base to the core.
 The results support evidence for acceleration of the jets with bulk motion Lorentz factor $\Gamma\propto R^{0.52\pm0.03}$ on de-projected scales $R$ of 0.5--500 parsecs. 
\end{abstract}



\section{Introduction}
\label{sec:intro}

The jets of active galactic nuclei (AGN) are believed to consist of relativistic electron-positron or electron-proton plasma moving in magnetic field. The emission mechanisms are synchrotron and inverse Compton, covering a spectrum range from radio up to $\gamma$-rays of TeV energies~\citep[e.g.,][]{1974ApJ...188..353J, 1980ApJ...235..386M}. Radio loud AGN show different pc-kpc morphology and properties depending on the orientation of their jets to the line of sight~\citep{1995PASP..107..803U,r:gaia4}. The smaller the viewing angle of a jet, the stronger is the Doppler boosting of the source emission leading to a ``favoritism'' in the observed properties of a flux limited AGN sample. Recent high-resolution observations of the jets in M\,87 and 1H\,0323$+$342 indicate the collimation and acceleration of the jets on scales up to hundreds of parsecs~\citep{2012ApJ...745L..28A, 2018ApJ...860..141H}. Acceleration on these scales is also seen with the very long baseline interferometry (VLBI) by measuring kinematics of bright knots in the jets~\citep{2009ApJ...706.1253H, 2015ApJ...798..134H, 2017ApJ...846...98J}.

The base of a jet seen with VLBI observations at cm-wavelengths corresponds to the unit optical depth surface (photosphere), called the ``core''. The apparent position and size of the core depend on the observation frequency~\citep{1979ApJ...232...34B}. This ``core shift`` effect can be inferred from multi-frequency VLBI experiments \citep[e.g.,][]{1984ApJ...276...56M,r:cs2008, 2009MNRAS.400...26O,2011A&A...532A..38S, 2012A&A...545A.113P,2013A&A...557A.105F,2014MNRAS.437.3396K,2016MNRAS.462.2747K,2017MNRAS.468.4478L,r:voitsik2018}. Not only do these measurements provide an information about the physical conditions in a jet, but also have an important practical implications for high-precision astrometry \citep[e.g.,][]{r:cs2008,2009A&A...505L...1P,r:gaia1,r:gaia2,r:gaia3,r:gaia4,r:gaia5}. 

At VLBI scales most jets look self-similar across a wide range of frequencies (and hence, angular resolutions): a clumpy quasi-conical shape structure consisting of emission knots and widening with apex distance. 
The physical nature of these emission knots remains unclear \citep{1995PNAS...9211348Z}, while it is widely assumed that these regions of enhanced emission are somehow associated with shock waves in the relativistic plasma flow \citep[e.g.,][]{1979ApJ...232...34B,1985ApJ...298..114M}.
Most of the knots move in a jet direction with super- or sub-luminal apparent speeds, while stationary components also seem to be a common occurrence \citep{2015A&A...578A.123R,2017ApJ...846...98J}, especially in BL~Lac type objects \citep{2018ApJ...853...68P}. A wide range of component speeds is often observed within a single source, which makes their relation to the velocity of the underlying plasma flow unclear~\citep[e.g.,][and references therein]{2016AJ....152...12L}.
The jet flow speed being considerably faster than the observed jet pattern speed is one of the possible explanations for high brightness temperatures measured by RadioAstron in some AGN \citep{2015A&A...583A.100L,r:RA_3C273}, while alternative interpretations of this phenomena might exist \citep{2002PASA...19...77K}.

In total flux density, the radio outbursts of AGN at lower frequencies are lagged with respect to that at higher frequencies. This is consistent with the opacity driven nature of the delays \citep[e.g.,][]{2016Galax...4...37M}.
In this scenario the peak of a flare at a given frequency might correspond to the moment when a disturbance traveling down the jet crosses the VLBI core at that frequency~\citep{2006A&A...456..105B, 2011MNRAS.415.1631K}.
The agreement between time lags and core shifts frequency dependencies found using simultaneous multi-frequency measurements gives a good support for that scenario~\citep{2014MNRAS.437.3396K}.
Within this assumption, one can estimate the speed of a moving disturbance in the core region using the peak-to-peak time delay and the core shift measured at two frequencies. Hereafter we refer to this jet speed unless the details are specified. This speed can be considered as a proxy to the plasma flow speed (or bulk motion speed) whether the flares are caused by a moving blob or a shock wave~\citep{2006AIPC..856....1M}. We note, however, that it might differ from the flow speed in a steady jet.
 
On the one hand the region probed with this method is well upstream the jet than that traced with VLBI kinematics analysis. On the other hand, there is an evidence that the major total flux density radio outbursts in AGN precede the occurrence of newborn VLBI components~\citep[e.g.,][]{2003ASPC..299..259S}. 
Therefore, a comparison of the speed in the core region with the pattern velocity seen by VLBI appears to be of a particular interest.


Through this paper we assume flat $\Lambda$CDM cosmology model with $H_0=69.3$\,km/(s\,Mpc), $\Omega_M=0.286$~\citep{2013ApJS..208...19H}, implemented by the \texttt{Astropy Python} library~\citep{2018arXiv180102634T}. We use positively defined spectral index $\alpha=d\ln S/d\ln\nu$.


\section{The sample and the data}
\label{sec:obsdata}

We selected radio-loud AGN that demonstrate exceptional variability in radio band and for which we were able to find the required multifrequency total flux density and VLBI data.
The sample includes two radio galaxies, two BL\,Lac type objects (hereafter BL\,Lacs), and seven flat spectrum radio quasars (hereafter quasars) according to their optical classification (Table~\ref{tab:src_params}). 
The prominent outbursts in their total flux density variations allow us to locate the peaks and estimate the multi-frequency peak-to-peak time delays. 
The optical class, redshift, and the observed median/maximal proper motion of the sources listed in Table~\ref{tab:src_params} are adopted from the MOJAVE kinematics paper~\citep{2016AJ....152...12L}.

The single-dish flux density monitoring observations of the objects were obtained with the 26\,m radio telescope of the University of Michigan Radio Observatory (UMRAO) at 8.0 and 14.5\,GHz. The sources are included in the UMRAO ``core sample'' and are monitored regularly over 40 years \citep[e.g.,][]{2017Galax...5...75A}. 

We analyzed the seven-frequency (4.6--43.2\,GHz) VLBA observations of the BL\,Lac type object 0851$+$202 (OJ\,287) and the high-redshift flat spectrum radio quasar 1633$+$382 (4C\,38.41). The observations of the two sources were performed in the framework of our survey of $\gamma$-ray loud blazars \citep{2010arXiv1006.3084S,2010arXiv1001.2591S} on 2009-02-02 (VLBA experiment BK150F) and 2009-06-20 (BK150L), respectively.
The quasar 2251$+$158 (3C\,454.3) investigated earlier by \cite{2014MNRAS.437.3396K} was also observed within this survey (BK150C, 2008-10-02).
The source 0851$+$202 (1633$+$382) was observed for 9 (13) hours during which the VLBA was switching between 
$C$ (6\,cm, 4.6/5.0\,GHz), $X$ (4\,cm, 8.1/8.4\,GHz), $K_u$ (2\,cm, 15.4\,GHz), $K$ (1\,cm, 23.8\,GHz), and $Q$ (7\,mm, 43.2\,GHz) band receivers. 
The $C$ and $X$ band data were split in two sub-bands centered at 4.6/5.0\,GHz and 8.1/8.4\,GHz, respectively. These sub-bands were independently analyzed.
The data were recorded with the legacy VLBA backend at the aggregate bitrate of 256\,Mbit/sec and correlated with the hardware VLBA correlator at the Array Operation Center in Socorro, NM,
USA.
The data reduction strategy follows the standard path of VLBA post-correlation analysis and fringe-fitting in \texttt{AIPS} \citep{2003ASSL..285..109G} and hybrid imaging \citep[e.g.][]{1995ASPC...82..247W} in \texttt{Difmap} \citep{1994BAAS...26..987S,1997ASPC..125...77S}. The ``preliminary imaging'' procedure \citep[used also by][]{2011A&A...532A..38S,2016MNRAS.462.2747K,2017MNRAS.468.4478L,2018arXiv180806138P} was employed to improve the amplitude calibration of the array to better than 5~percent at frequencies below 22\,GHz (better than 10~percent at 23.8 and 43\,GHz). For additional details of the performed VLBA data reduction see \cite{2011PhDT.........6S}.

\begin{table}
\caption{Sources and their parameters. Columns are: (1) B\,1950 source name; (2) redshift; (3) -- source optical class (Q -- quasar, B -- BL\,Lac object and G -- radio galaxy); (4--5) median and maximal VLBI apparent proper motion as measured by MOJAVE; (6) luminosity distance.}
\label{tab:src_params}
\begin{tabular}{ccp{1em}ccc}
\hline
\hline
Source  &  z & OC &   $\umu_\mathrm{app, med}^\mathrm{M}$ &   $\umu_\mathrm{app, max}^\mathrm{M}$ &  $D_L$ \\
		&          & 	    &    (mas/yr) &  (mas/yr) &  (Mpc) \\
(1) &  (2) &  (3) &  (4) &  (5) & (6) \\
\hline
0415$+$379 & 0.0491 & G & $1.47 \pm 0.33$ & $2.51 \pm 0.10$ & 220  \\
0420$-$014 & 0.9161 & Q & $0.08 \pm 0.03$ & $0.12 \pm 0.01$ & 6033 \\
0430$+$052 & 0.0330 & G & $2.21 \pm 0.39$ & $2.94 \pm 0.13$ & 146  \\
0607$-$157 & 0.3226 & Q & $0.04 \pm 0.02$ & $0.06 \pm 0.04$ & 1711 \\
0851$+$202 & 0.3060 & B & $0.34 \pm 0.30$ & $0.79 \pm 0.02$ & 1609 \\
1308$+$326 & 0.9970 & Q & $0.35 \pm 0.14$ & $0.53 \pm 0.02$ & 6701 \\
1633$+$382 & 1.8140 & Q & $0.15 \pm 0.16$ & $0.37 \pm 0.02$ & 14078 \\
1730$-$130 & 0.9020 & Q & $0.31 \pm 0.12$ & $0.57 \pm 0.03$ & 5918 \\
2200$+$420 & 0.0686 & B & $1.08 \pm 0.55$ & $2.21 \pm 0.16$ & 312  \\
2223$-$052 & 1.4040 & Q & $0.21 \pm 0.04$ & $0.27 \pm 0.05$ & 10253 \\
2251$+$158 & 0.8590 & Q & $0.20 \pm 0.16$ & $0.30 \pm 0.01$ & 5571  \\
\hline
\end{tabular}
\end{table}


\section{Light curves analysis}
\label{sec:lc}

\subsection{Gaussian process regression}
\label{sec:gpr}

To obtain the time of a flare peak at each frequency we employ Gaussian process regression~\citep[GPR, see][]{Rasmussen:GPM}.
This Bayesian machine-learning 
technique does not require any prior information about the signal and provides a probabilistic distribution of the data at any given point. GPR was recently used for the analysis of blazars' light curves \citep{2016A&A...590A..48K} and provides time lag measurements consistent with the widely-used discrete correlation function \citep{1988ApJ...333..646E} analysis, but with smaller uncertainties \citep{r:RA_0235} when applied to the same light curves. 
The Gaussian process (GP) is characterized by the covariance function (\textit{kernel}). The kernel function is expressed in terms of \textit{hyperparameters} which are learned from the data by maximizing the marginal likelihood function (\textit{training} the GP). 

\begin{figure}
	\includegraphics[width=\columnwidth, trim=0.5cm 0.8cm 0.3cm 0.4cm]{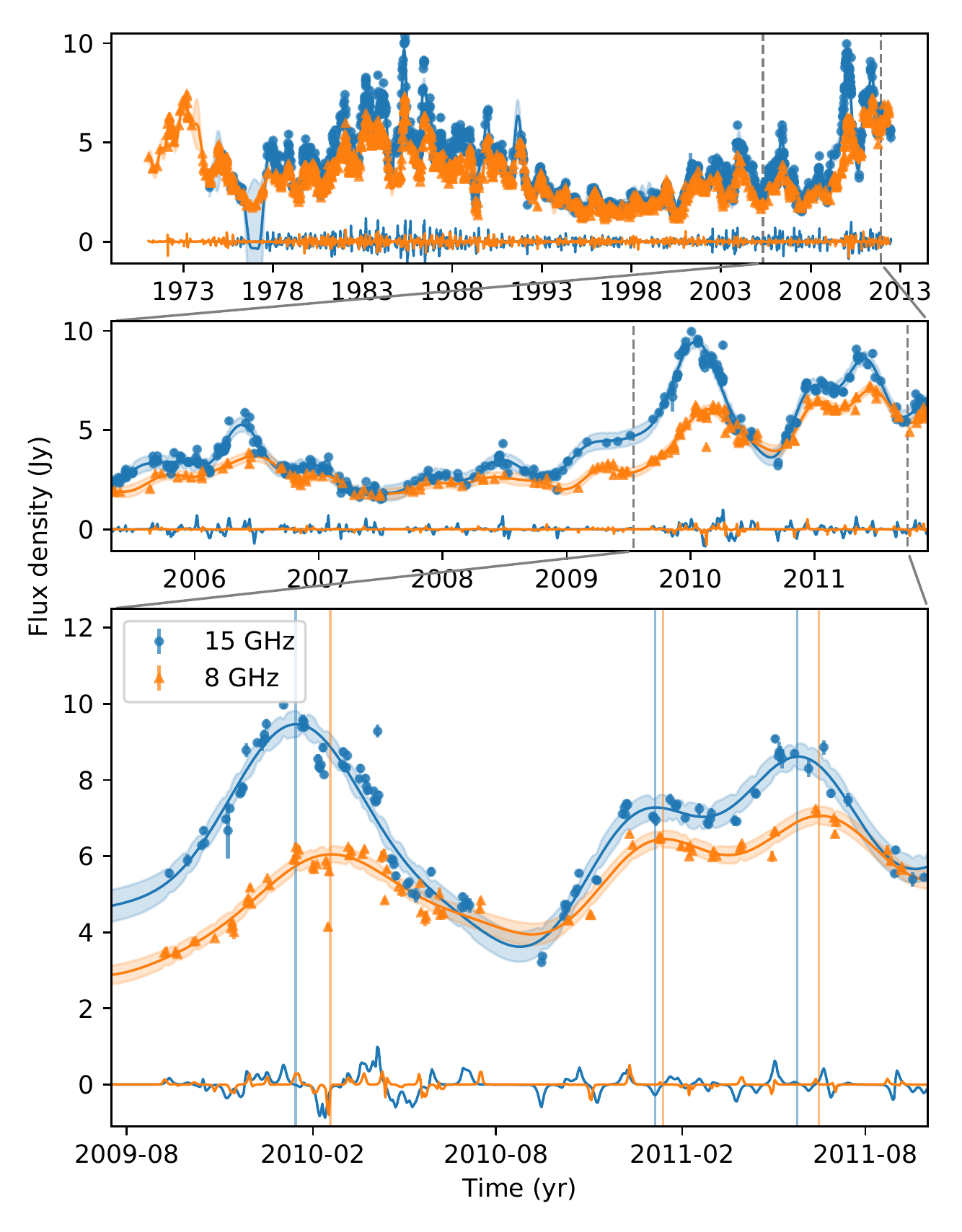}
	\caption{The light curves of 0851+202 (OJ\,287) at 15 and 8\,GHz. The SE and MT (with the zero-mean) components of the GPR fit are shown separately with the curves. The $\pm\sigma$ confidence bounds are shown around the SE-component as a shaded area. The dashed rectangle on the panels denotes the region zoomed. The vertical lines in the bottom panel show the peaks automatically-detected in 15 and 8\,GHz light curves (see Section~\ref{sec:lc}).}
	\label{fig:lc0851}
\end{figure}

\begin{figure}
	\includegraphics[width=\columnwidth, trim=0.3cm 1cm 0.3cm 0.3cm]{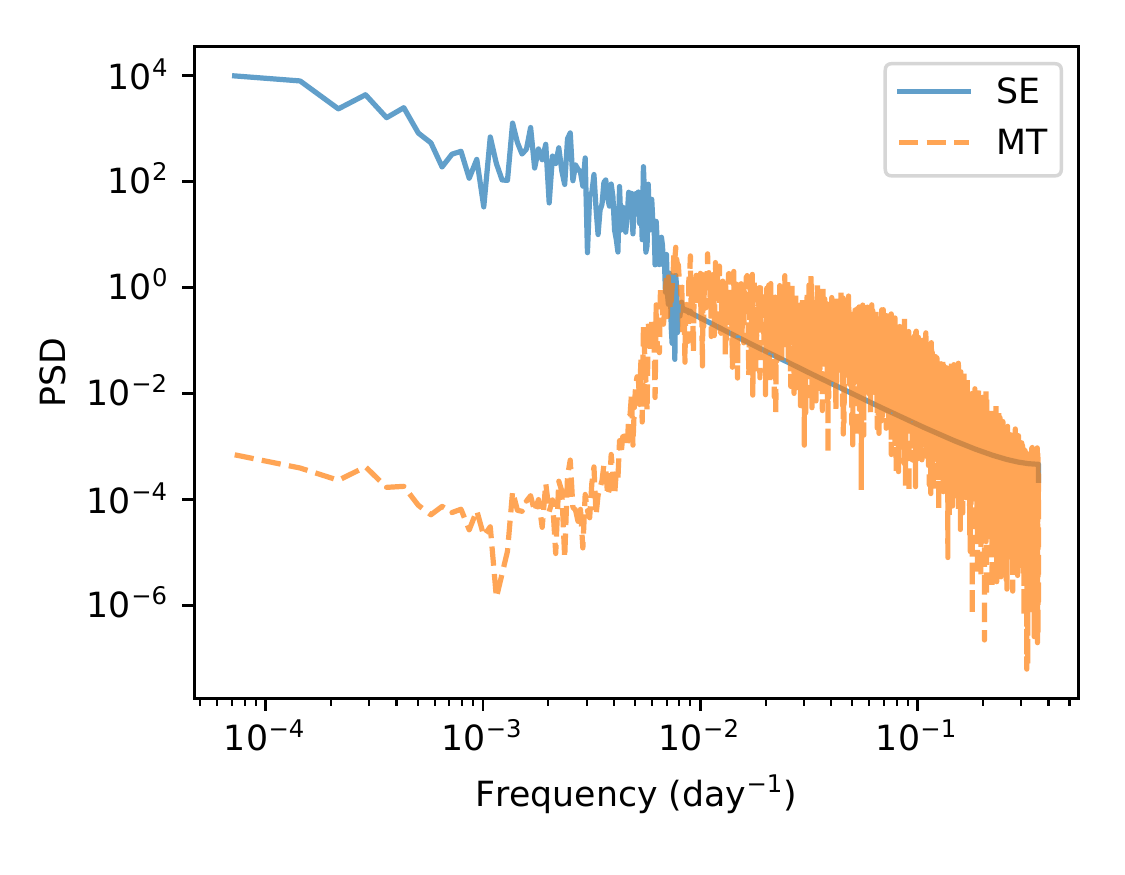}
	\caption{Power spectra of the long-term (SE) and short-term (MT) light curve regression components (OJ\,287, 15\,GHz).}
	\label{fig:psd}
\end{figure}

Total flux density variations in our sample AGNs have a non-zero power on timescales from days to years (e.g. Fig~\ref{fig:lc0851}) showing a ``colored'' power spectra (Fig.~\ref{fig:psd}). 
At 8\,GHz the variations are much smother than that at 15\,GHz. 
This leads to an ambiguity in cross-identification of the flare peaks. 
We use the kernel represented by a mixture of \textit{Squared Exponential} (SE) and \textit{Mat\'{e}rn} (MT) kernels to 
model the long- (``flares'') and short-term (``flickering'') variations, respectively.
We use the same metric scale of SE-kernel inferred from the GPR optimization for both 5 and 8\,GHz lightcurves. 
This helps us to cross-identify major flares between the two frequencies. 
The other hyperparameters (SE-amplitude, MT-metrics and MT-amplitude) are optimized independently for 5 and 8\,GHz data. 

Once the GP is optimized, one can infer its Bayesian prediction and confidence bounds for a given kernel. 
In Figure~\ref{fig:lc0851} the light curves of OJ\,287 at 15 and 8\,GHz are shown at various zoom levels. The SE and MT components of the GPR fit are shown separately (the latter has zero mean value). The confidence bounds of $\pm\sigma$ are shown with a shaded area around the SE-component and estimated as $\sigma = \sqrt{\sigma_{\mathrm{SE}} + \sigma_{\mathrm{MT}}}$, where $\sigma_{\mathrm{SE}}$ and $\sigma_{\mathrm{MT}}$ are the values obtained from the covariance matrices of GP prediction for SE and MT kernels, correspondingly. 

In Figure~\ref{fig:psd} the power spectra of these two components are shown. Their sum gives the ``red-noise'' power spectrum typical for blazars~\citep[e.g.,][]{2014MNRAS.445..437M}, with a slope of about $-2$. 
The SE-component fits flux density variations longer than about 100 days, while the MT one deals with a short term variability. 
The MT-component not only describes a white noise but also models the correlated fast variations in the data, as can be seen from its non-flat power spectrum. These variations become more prominent during the source flaring state. Their amplitude is higher at higher frequency. During the major flares these variations correlate between each other at 15 and 8\,GHz with almost a zero lag, while outside the flares they are uncorrelated. If cross-correlation technique or a single-kernel GPR is used, this flickering will prevent an accurate peak detection. The nature of these fast-term variations might be a subject for a separate study.

The combination of SE and MT kernels is a good choice for cross-identification of the flares at different observing frequencies, but if one is interested in describing a single light curve fine structure, the \textit{Rational Quadratic} kernel would be a good choice as well~\citep{r:RA_0235}. For the GPR we employed the \texttt{george} python library \citep{2015ITPAM..38..252A}.

\subsection{Flare detection and variability timescale}
\label{sec:flares}

Having the GPR prediction on the SE-component $S(t)$ we use the following algorithm for automatic flares detection. A peak position $t_{\mathrm{max}}$ is adopted where the gradient (the first derivative) of GPR changes its sign, the second derivative is negative $\ddot{S}<0$, and the peak exceeds a $2\sigma$ threshold. In the last panel of Figure~\ref{fig:lc0851} the automatically-detected peaks in 15 and 8\,GHz light curves are shown with vertical lines.  

We located the inflection points $t_{\mathrm{left}}$ and $t_{\mathrm{right}}$ just left and right of the $t_{\mathrm{max}}$ (where the second derivative changes sign).
These measurements are summarized in Table~\ref{tab:flares}.
Around each inflection point $t_i$ we estimate the ``$e$-folding'' timescale as $\tau_i = \sqrt{2(t_{i+1}-t_{i-1})} / \ln (S_{i+1}/S_{i-1})$ (since a GPR realization implies $S(\Delta t)=Ae^{\Delta t^2/\tau^2}$). The shortest time scale attained at the inflection points provides the lowest estimate of the size of the region where the variability originates. 

After all the flares are detected, their cross-identification is performed. In addition to the automatic determination of the close peaks we manually inspected their similarity for the final conclusion. For example, in the light curves of OJ\,287 we successfully cross-identified 18 flares, which is the highest number among all the sources. We estimate an error of $t_{\mathrm{max}}$ for each flare from a distribution of peaks of 100 GP samples taken between $t_{\mathrm{left}}$ and $t_{\mathrm{right}}$. An error of the variability scale is estimated in the same way.
For each source, we calculate the weighted averaged time delay $\Delta t$ and the time scale $\tau$. The results are listed in Table~\ref{tab:res}. 

\begin{table*}
\caption{Measured parameters of the cross-identified flares at 15 and 8\,GHz.}
\begin{tabular}{c|c|c|c|c|c|c|c|c|c|c|c|c}
\hline\hline
Source & $t_\mathrm{15, max}$ & $t_\mathrm{15, err}$ & $f_\mathrm{15, max}$ & $f_\mathrm{15, err}$ & $t_\mathrm{15, left}$ & $t_\mathrm{15, right}$ & $t_\mathrm{8, max}$ & $t_\mathrm{8, err}$ & $f_\mathrm{8, max}$ & $f_\mathrm{8, err}$ & $t_\mathrm{8, left}$ & $t_\mathrm{8, right}$ \\
 & MJD & days & Jy & Jy & MJD & MJD & MJD & days & Jy & Jy & MJD & MJD \\
(1) & (2) & (3) & (4) & (5) & (6) & (7) & (8) & (9) & (10) & (11) & (12) & (13) \\
\hline
0415$+$379 & 53404.9 & 7.7 & 4.28 & 0.22 & 53280.1 & 53503.7 & 53456.4 & 9.6 & 5.28 & 0.17 & 53317.1 & 53557.9 \\
 & 53863.1 & 6.7 & 5.35 & 0.22 & 53726.0 & 53971.6 & 53943.7 & 8.7 & 6.15 & 0.16 & 53793.4 & 54028.9 \\
 & 54491.6 & 1.9 & 8.57 & 0.21 & 54349.9 & 54611.1 & 54596.3 & 5.8 & 8.54 & 0.17 & 54446.7 & 54696.5 \\
 & 54911.4 & 5.8 & 6.52 & 0.21 & 54828.2 & 55059.6 & 54973.7 & 7.7 & 7.91 & 0.09 & 54893.0 & 55126.0 \\
0420$-$014 & 44133.8 & 30.2 & 6.14 & 0.3 & 43990.0 & 44385.8 & 44161.2 & 8.5 & 5.37 & 0.14 & 44020.2 & 44308.5 \\
 & 44782.9 & 7.5 & 5.05 & 0.31 & 44660.7 & 44894.9 & 44910.7 & 13.2 & 4.51 & 0.19 & 44756.3 & 45006.4 \\
 & 45467.6 & 9.4 & 5.31 & 0.31 & 45353.1 & 45608.9 & 45493.8 & 11.3 & 4.9 & 0.18 & 45376.4 & 45643.0 \\
 & 50008.7 & 24.5 & 4.76 & 0.04 & 49756.7 & 50329.5 & 50207.3 & 25.5 & 3.64 & 0.2 & 50054.9 & 50349.6 \\
\hline
\end{tabular}
\label{tab:flares}
\begin{flushleft}
The full table is available online in a machine readable form. \textbf{Column designation:} (1)~--~Source name; (2)~--~Peak time of a flare at 15\,GHz; (3)~--~Error in the peak position; (4--5)~--~15\,GHz peak flux density and the error; (6--7)~--~15\,GHz inflection points positions; (8--13)~--~The same parameter as in columns 2--7 for 8\,GHz light curves.
\end{flushleft}
\end{table*}


\section{Core shift measurements}
\label{sec:cs}

As noted before, the core shift amplitude can change within a source between the epochs of measurements. In this study we are interested in the mean amplitude of the effect. Therefore, we use all the available core shifts estimates from the literature as well as perform our own measurements for the two sources in our sample.

We measure the 15.4--8.1\,GHz core shifts in the blazars 0851$+$202 and 1633$+$382 directly using VLBA observations (Section~\ref{sec:obsdata}).
A relative shift between the apparent core position observed at two frequencies is obtained as the difference of the shift between the images and the offset between the core position at each frequency map~\citep[e.g.,][]{2012A&A...545A.113P}. The image shift vector is measured by aligning the cross-identified optically thin jet components. The offset vector is derived from the models as the difference of the core radius-vectors. The position errors of the components are estimated in the image plane following~\citet{1999ASPC..180..301F} and then propagated to the core shift estimates.

\begin{figure}
	\includegraphics[width=\columnwidth, trim=0.3cm 0cm 0.3cm 0.5cm]{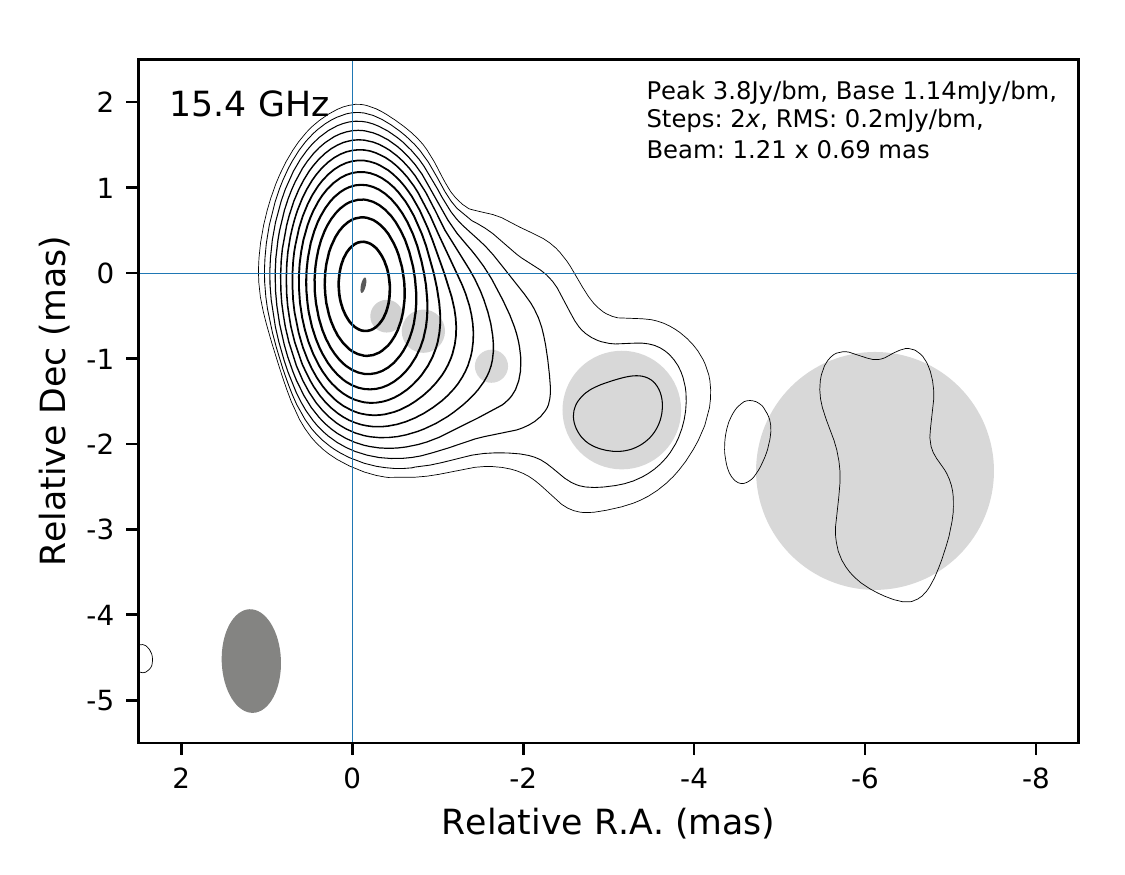}	\includegraphics[width=\columnwidth, trim=0.3cm 0.4cm 0.3cm 0.6cm,clip]{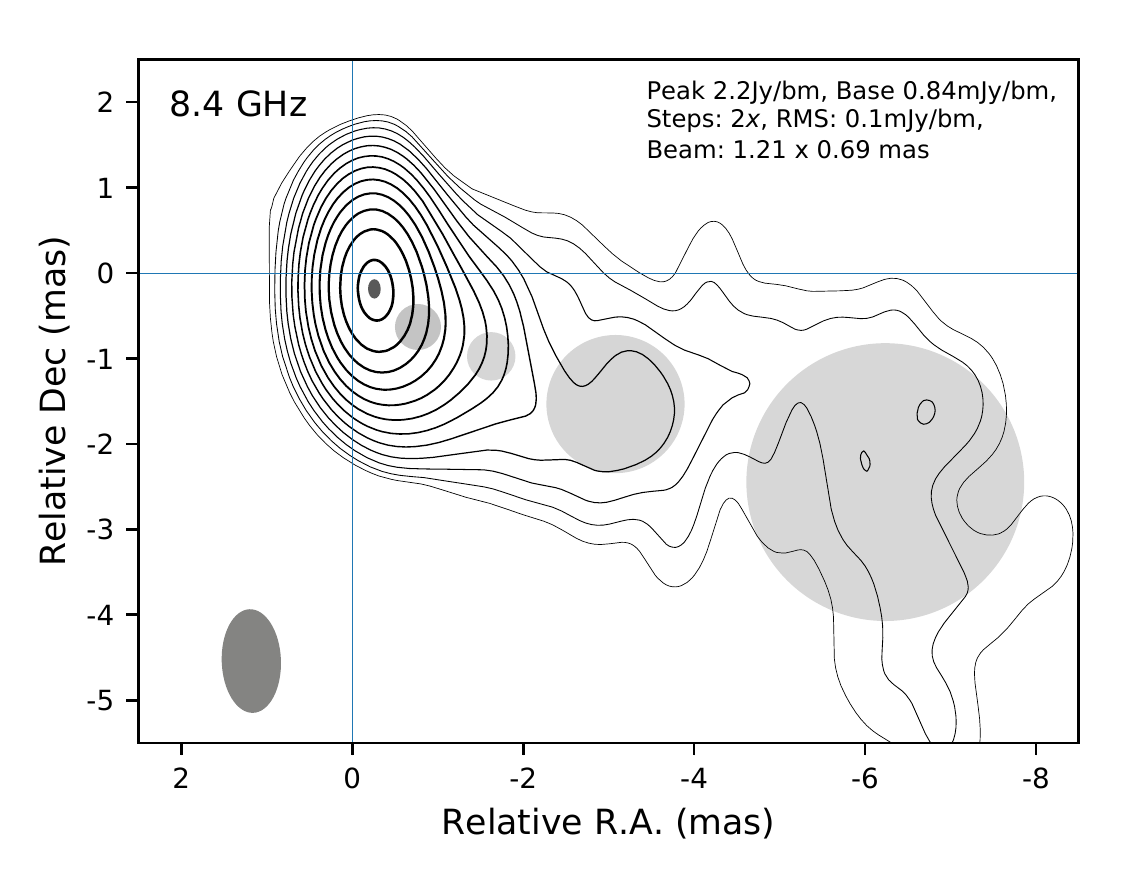}
	\caption{Naturally weighted \texttt{CLEAN} maps of OJ\,287 at 15 and 8\,GHz. The beam is shown in the lower left corner. The core is modeled with an elliptical Gaussian component. The rest model components are shown with shaded circles.}
	\label{fig:cmaps}
\end{figure}

As an example, in Figure~\ref{fig:cmaps} the \texttt{CLEAN} maps of OJ\,287 at 15 and 8\,GHz are shown. The images are shown with the same beam corresponding to that at the lower frequency for a better visualization. The structure of the sources was modeled in the $uv$-plane using \texttt{Difmap}. 

For the blazar 2251$+$158 we use the core shift measurements by \citet{2014MNRAS.437.3396K}. For BL\,Lacertae (2200$+$420) we use the results by~\citet{2009MNRAS.400...26O} $\Delta r_{15-8}=0.18\pm0.04$ mas. 
We also use the 15--8\,GHz core shift measurements by~\citep[][Table 1]{2012A&A...545A.113P}. When several core shift measurements are available for an object (four cases), they are averaged, and the conservative maximal error is adopted as the uncertainty. 
The averaged core shifts used for the analysis are listed in column (3) of Table~\ref{tab:res}.

\section{Results and discussion}
\label{sec:results}

\begin{table*}
	\begin{minipage}{0.79\linewidth}
		\centering{
			\caption{Measured core shifts, time lags, and estimated parameters of the AGN}
			\label{tab:res}	
			\begin{tabular}{@{}c@{~}c@{~}c@{~}c@{~}c@{~~}c@{~~}c@{~~}c@{~~}c@{~~}c@{~~}c@{~~}c@{~~}c}
				\hline 
				\hline 
				Source & Name & $N$ & $\Delta r$ & $\Delta t$ & $\umu_\mathrm{app}$ & $\beta_{\mathrm{app}}$ & $\tau$ & $\delta$ & $\Gamma$ & $\theta$ &
				$R_{15}$ & $R_8$ \\
				& &	&	mas			& days &			mas/yr	& $c$	&	days & & & deg & pc & pc \\
				(1)   &	(2) &	(3) & (4) &			(5)	& (6)	&	(7) & (8) & (9) & (10) & (11) & (12) & (13) \\	   
				\hline
0415$+$379 & 3C\,111 & 5 & 0.315 & $89.9\pm4.2$ & $1.2\pm0.2$ & $4.0\pm0.7$ & 46.2 & 5.3 & 4.3 & 10.5 & 1.9 & 3.6\\
0420$-$014 &  & 8 & 0.267 & $63.5\pm3.8$ & $0.8\pm0.2$ & $39.9\pm7.8$ & 47.9 & 70.2 & 46.4 & 0.7 & 198.7 & 372.6\\
0430$+$052 & 3C\,120 & 7 & 0.075 & $13.0\pm2.4$ & $2.0\pm1.4$ & $4.6\pm3.2$ & 36.1 & 8.1 & 5.4 & 6.1 & 0.5 & 1.0\\
0607$-$157 &  & 7 & 0.240 & $50.9\pm4.4$ & $1.3\pm0.3$ & $26.6\pm6.0$ & 44.9 & 44.2 & 30.1 & 1.1 & 65.1 & 122.1\\
0851$+$202 & OJ\,287 & 18 & 0.116 & $20.5\pm1.5$ & $1.6\pm0.4$ & $30.9\pm8.3$ & 26.7 & 32.7 & 30.9 & 1.7 & 20.0 & 37.4\\
1308$+$326 &  & 8 & 0.143 & $40.1\pm4.3$ & $0.7\pm0.2$ & $34.6\pm12.6$ & 40.7 & 61.5 & 40.5 & 0.8 & 95.8 & 179.7\\
1633$+$382 &  & 3 & 0.201 & $51.4\pm13.8$ & $0.5\pm0.2$ & $40.2\pm13.3$ & 73.2 & 50.3 & 41.2 & 1.1 & 102.4 & 192.0\\
1730$-$130 &  & 3 & 0.174 & $50.3\pm7.3$ & $0.7\pm0.2$ & $32.6\pm10.5$ & 52.2 & 47.6 & 35.0 & 1.1 & 80.5 & 151.0\\
2200$+$420 & BL\,Lac & 13 & 0.106 & $15.9\pm2.7$ & $2.3\pm1.0$ & $10.5\pm4.8$ & 25.7 & 6.5 & 11.8 & 7.9 & 1.2 & 2.2\\
2223$-$052 & 3C\,446 & 8 & 0.199 & $50.9\pm5.7$ & $0.6\pm0.2$ & $40.0\pm11.0$ & 54.7 & 80.8 & 50.3 & 0.6 & 198.8 & 372.8\\
2251$+$158 & 3C\,454.3 & 9 & 0.189 & $37.1\pm2.7$ & $1.0\pm0.2$ & $47.3\pm11.8$ & 44.2 & 110.5 & 65.4 & 0.4 & 257.2 & 482.3\\
				\hline
			\end{tabular}
		}
		\begin{flushleft}
			\textbf{Columns designation:} (1-2)~source B1950 and alternative names; (3)~number of cross-identified flares between 15 and 8\,GHz; (4)~mean 15-8\,GHz core shift; (5)~weighted averaged 15-8\,GHz time delay; (6)~apparent angular speed; (7)~apparent speed in units of speed of light (Eq.\,\ref{eq:beta_app}); (8)~weighted averaged  variability time scale of the flares at 15\,GHz; (9)~estimated Doppler factor; (10)~estimated Lorentz factor; (11)~estimated viewing angle; (12--13)~de-projected distance from the jet base to the core at 15 and 8\,GHz.
		\end{flushleft}
	\end{minipage}
\end{table*}

\subsection{Variability time scales}

\begin{figure}
	\includegraphics[width=\columnwidth, trim=0 1cm 0 0, trim=0 1cm 0 0]{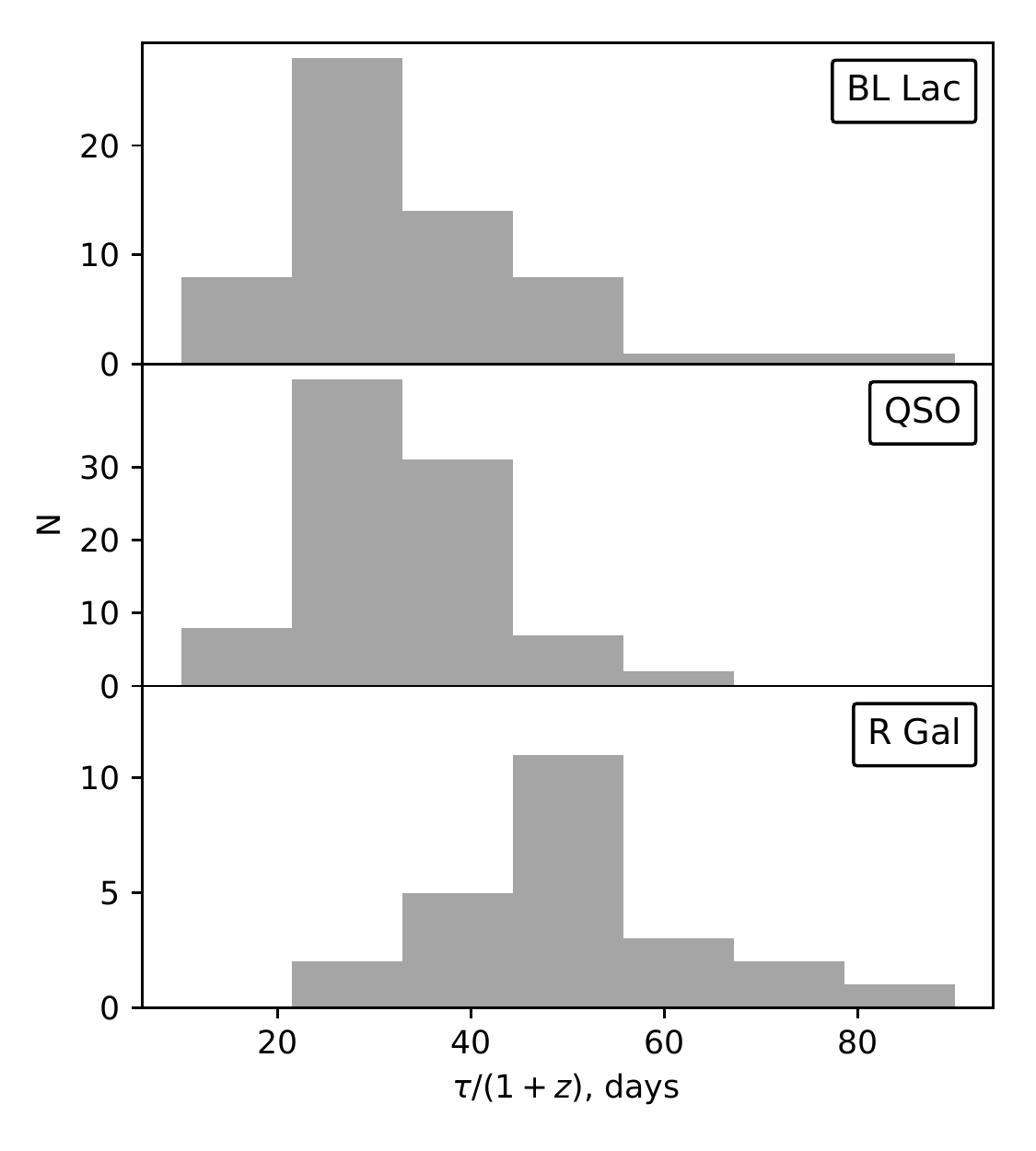}
	\caption{Variability time scale distributions at 15\,GHz for different object types over all the flares.}
	\label{fig:scales}	
\end{figure}

The variability time scales estimated in Sec.~\ref{sec:flares} vary from one flare to another within one source. The distribution of the time scales in the source frame is shown in Figure~\ref{fig:scales} for all flares. 
We compared the distributions using the two-sample Anderson-Darling test~\citep{anderson1952}. The test fails to reject the null hypothesis that the time scales in the BL\,Lacs and the quasars have the same distribution (with the $p$-value of 0.093). This is expected for the similar relation between their redshifts and Dopper factors (Section~\ref{sec:doppler}). The radio galaxies have significantly longer variability timescales which is explained by much lower Doppler factors in their core regions. We note, that the GPR approach is flexible enough to represent a variability over a wide range of timescales. Hence, we do not expect a bias either due to the method used or due to the data properties.

\subsection{Core shifts and time delays}
\label{sec:cs_dt}

\begin{figure}
	\includegraphics[width=\columnwidth, trim=0 1cm 0 0]{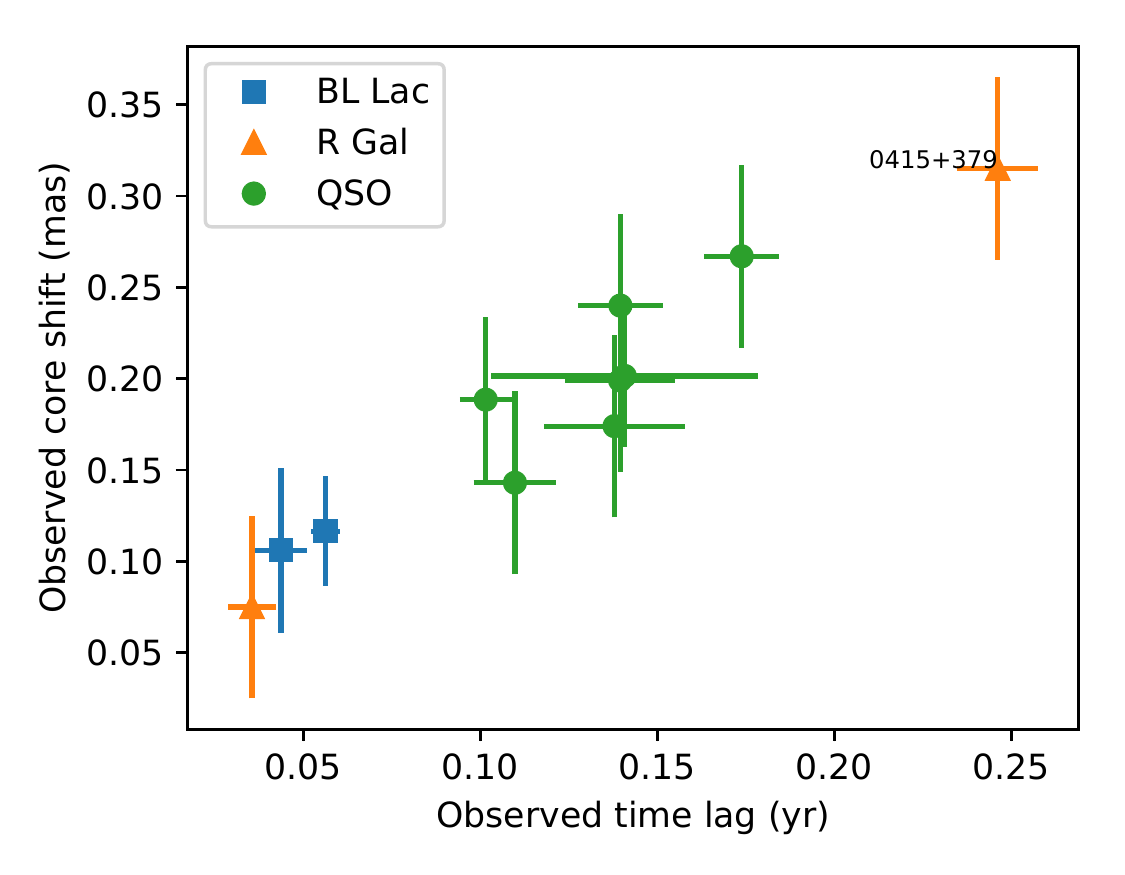}
	\caption{The observed averaged 15--8\,GHz core shifts and peak-to-peak time delays between 15 and 8\,GHz flares in the sources.}
	\label{fig:cs_dt_obs}
\end{figure}

A strong correlation between an observed averaged projected core shift and an observed peak-to-peak total flux density time delay is found in the sources, as shown in Figure~\ref{fig:cs_dt_obs}. 
The Spearman correlation coefficient is $S_c=0.97$ with the $p$-value of $10^{-7}$. The latter gives an estimate of the probability to obtain the above value of $S_c$ (or a more extreme one) for two uncorrelated samples. We perform a linear fit and use Markov chain Monte Carlo (MCMC) to obtain the posterior distributions of the parameters. The core shifts and time delays in our sample relate as 
$\Delta r[\mathrm{mas}] = (1.5\pm0.1)\Delta t[\mathrm{yr}]$. 

The reason for this tight correlation might be related to the sample selection (see below). We also highlight, that 
other powerful sources should have similar relation between the core shifts and time lags. For them, one can obtain an estimate of the core shift from the time lag measurements and vice versa. 

Thereby, for the first time we confirm a tight relation between the apparent core shifts and the observed flare delays in AGNs. 
The sources in the sample have redshifts ranging from $z=0.03$ to $1.8$. Neither the time lags nor the core shifts in our sample show correlation with the redshift. Moreover, the two radio galaxies (nearest to us) show minimal and maximal core shifts and time delays. We discuss the reasons for this correlation in Section~\ref{sec:accel}.

\subsection{The speeds}

Assuming that the time delay between flare peaks at 15 and 8\,GHz corresponds to the time it takes a disturbance to travel from the 15\,GHz core to the 8\,GHz core, one can estimate the apparent speed of the jet in this region as:

\begin{equation}\label{eq:beta_app}
\beta_\mathrm{app} = \frac{D_A\Delta r}{c \Delta t},
\end{equation}
where $D_A$ and $\Delta r$ are the angular diameter distance to the source and the apparent core shift. The apparent angular velocity is $\umu_\mathrm{app}=\beta_{\mathrm{app}}c/D_M = (1+z)^{-1}\Delta r/\Delta t$, where $D_M$ is the proper motion distance~\citep[e.g.,][]{1999astro.ph..5116H}. These estimates are summarized in Table~\ref{tab:res}. The errors are propagated from the corresponding uncertainties of $\Delta r$ and $\Delta t$ measurements. 


\begin{figure*}
	\includegraphics[width=\columnwidth, trim=0 0.5cm 0 0]{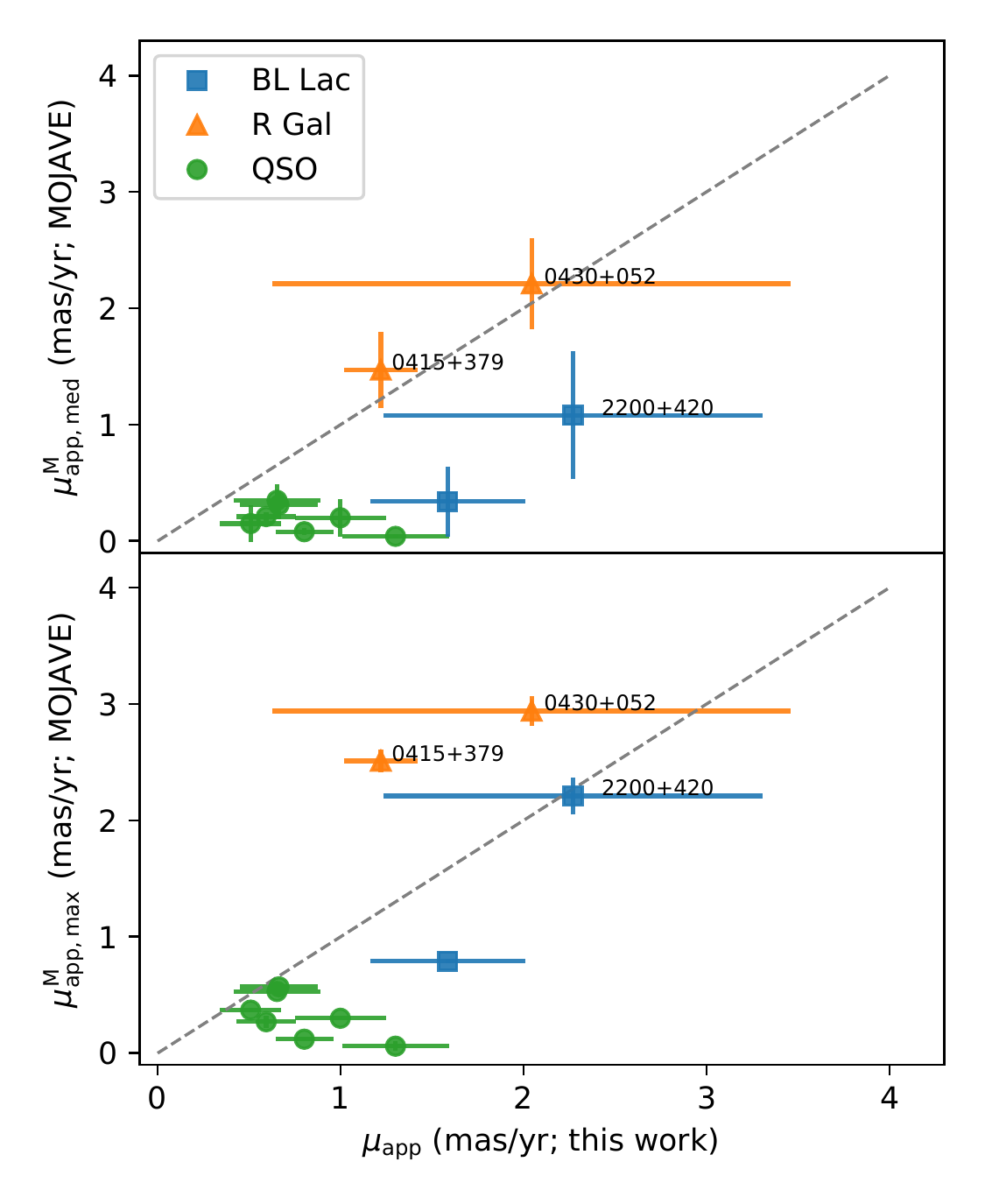}~\includegraphics[width=\columnwidth, trim=0 0.5cm 0 0]{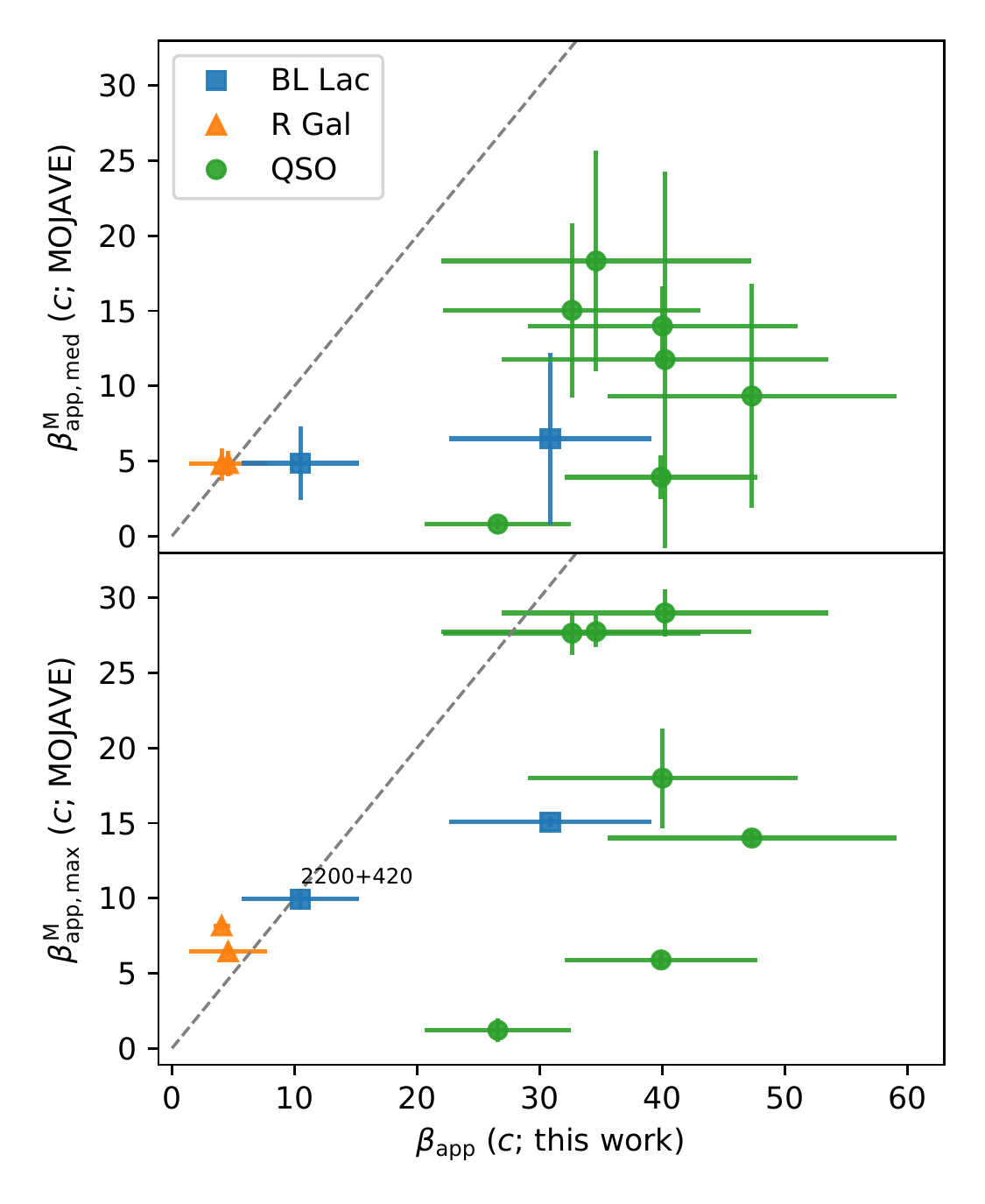}
	\caption{\textit{Left}: the median and maximal proper motion of the sources measured by MOJAVE vs. that obtained in this work. \textit{Right:} the same for apparent transverse speed in units of $c$. The dashed lines show equality.}
	\label{fig:speeds}	
\end{figure*}

In Figure~\ref{fig:speeds} the estimated apparent angular and linear speed is compared to that measured by~\citet{2016AJ....152...12L} for moving jet features. The former is higher than that inferred from component kinematics for all the quasars in our sample, while for the radio galaxies the situation is opposite. This can occur in an accelerating jet, as discussed in Section~\ref{sec:accel}. 

The capability of VLBI to catch the true jet speed is dictated by the source characteristics like the jet structure and Doppler factor, but also by the observations parameters like sensitivity, image fidelity, robustness of the model fitting and component cross-matching between epochs as well as the time coverage. These factors might prevent detection of the true plasma speed by VLBI kinematics analysis. 

\begin{figure}
	\includegraphics[width=\columnwidth, trim=0.3cm 0.8cm 0.3cm 0.2cm]{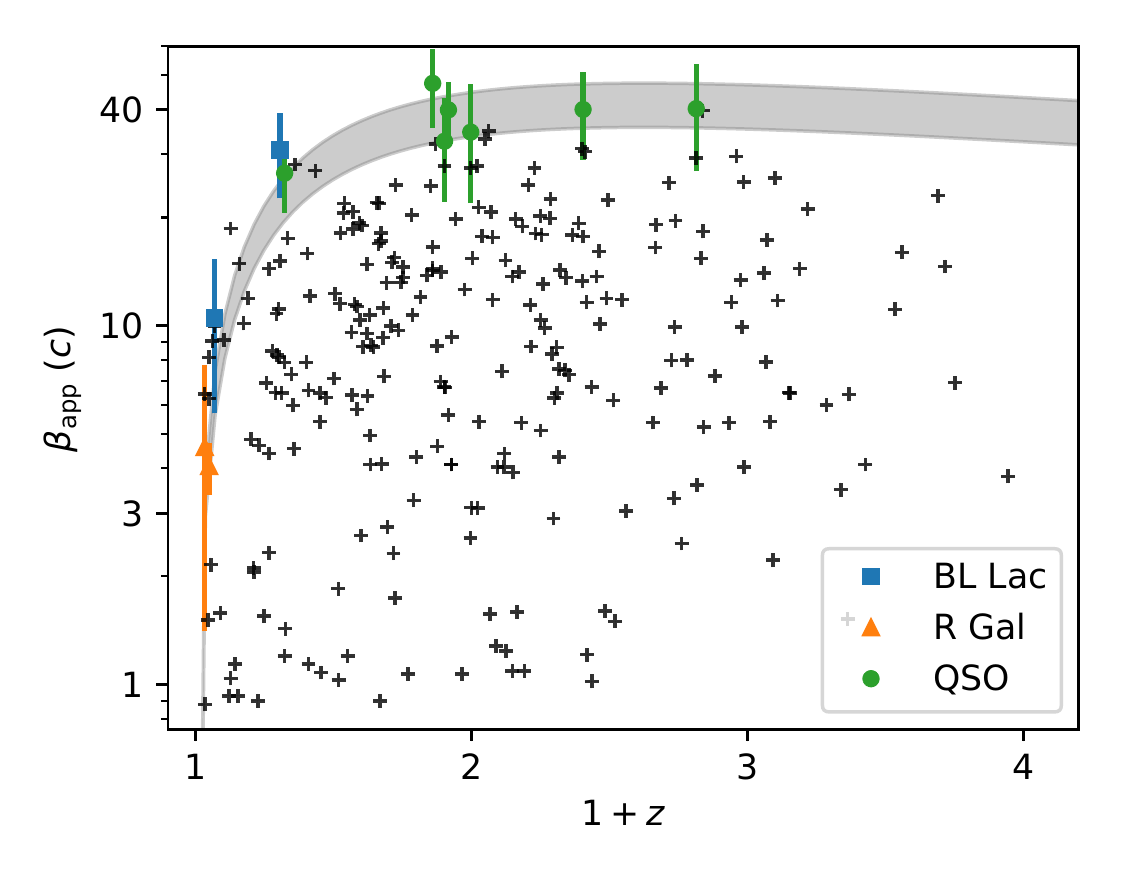}
	\caption{Maximal VLBI apparent speed vs. the redshift for the MOJAVE sample sources~\citep[black crosses,][]{2016AJ....152...12L}. Our estimates are shown with the barred symbols. The shaded area includes 95\% of the posterior MCMC samples obtained for the ``core shift -- time lag'' fit in Sec.~\ref{sec:cs_dt}}
	\label{fig:beta_z}
\end{figure}

In Figure~\ref{fig:beta_z} the apparent speed is plotted against the redshift. Our results are consistent with the upper envelope of $\beta_\mathrm{app, max}$ measured by VLBI among all the sources in MOJAVE sample. These values are plotted with black crosses~\citep[see Fig.~8 by][]{2016AJ....152...12L}. The shaded area includes 95\% of posterior MCMC samples obtained for $\Delta r/\Delta t$ data (Fig.~\ref{fig:cs_dt_obs}). One can see many distant objects falling into this interval, which implies that the relation between a core shift and a time delay holds for them. Below we provide an explanation of why the estimated apparent speeds in the jets follow this trend. For this purpose we estimate more parameters of the sources.

To estimate the speeds we used the core shift values measured irregardless a source flaring state, implying that the most were obtained for an undisturbed jet. It is known, however, that during the major flares the core shift increases, presumably due to the raise of particles density~\citep[e.g.,][]{2017MNRAS.468.4478L}. As seen in Figure~\ref{fig:lc0851}, the time delays are much shorter than the overall flare duration at 8 and 15\,GHz, i.e. the core is disturbed at both frequencies when the time lag is measured. This might introduce a bias resulting in the underestimation of the speeds.

Another possible source of a bias can be caused by a presence of a stationary component in the jet. It can affect both the core shift (by pulling on the brightness) and the time delay (causing additional substructure of a flare) measurements. Both would be underestimated if a stationary feature resides between the cores at 15 and 8\,GHz, and overestimated otherwise. The further such a feature is from the core, the smaller its effect on the measurements would be. In case when it dominates the emission, one would expect non $1/\nu$ core shift dependence.

\subsection{Doppler factors, Lorentz factors and viewing angles}
\label{sec:doppler}

Assuming that the variability timescale corresponds to the core light-crossing time, we estimated the Doppler factor using the timescale $\tau$ derived above and the average 15\,GHz core size $a$ derived from MOJAVE VLBI observations (FWHM of the circular Gaussian core model) by \cite{2016AJ....152...12L}. Since we deal with the flares, the core size is averaged with the weights proportional to its flux density. 
According to \cite{2005AJ....130.1418J,2017ApJ...846...98J}:
\begin{equation}\label{eq:dvar}
\delta = \frac{25.3\,\,(a/\mathrm{mas}) \,\,(D_L/\mathrm{Gpc})}{(\tau/\mathrm{yr})\,\,(1+z)}.
\end{equation}

Further we estimate the Lorentz factors and the viewing angles of the jets using the following equations:

\begin{equation}\label{eq:gamma}
\Gamma = \frac{\beta_\mathrm{app}^2 + \delta^2 + 1}{2\delta},\;\;\;\;\;
\theta = \arctan \frac{2\beta_\mathrm{app}}{\beta_\mathrm{app}^2 + \delta^2 - 1}.
\end{equation}

These results are summarized in Table~\ref{tab:res}. Our estimates of Doppler factors are consistent with that performed by~\citet{2009A&A...494..527H} for the two radio galaxies and for BL\,Lacertae, but are several times higher for the rest of the objects. The discrepancy is caused by different apparent speeds used in the calculations. Moreover, the assumption of the constant intrinsic brightness temperature equal to the equipartition value \citep{1994ApJ...426...51R} made by the authors might be violated in those sources \citep{r:RA_BL,r:RA_0529,r:RA_0235}.

The derived Lorentz factor values range from 4 in radio galaxies to about 40 in quasars. The highest value $\Gamma=65$ is obtained for the blazar 3C\,454.3. Such extreme value of $\Gamma$ is consistent with naive expectations for the ``most extreme blazar'' as 3C\,454.3 is the brightest flaring $\gamma$-ray blazar observed so far \citep{2011ApJ...733L..26A,2011ApJ...736L..38V} while $\gamma$-ray bright blazars are expected to have higher Doppler boosting factors (\citealt{2009ApJ...696L..22L,2011ApJ...726...16L}, but not necessary higher values of $\Gamma$, \citealt{2010A&A...512A..24S}).
However, the observed spectral energy distribution of 3C\,454.3 may be modeled without requiring such high values of $\Gamma$ and $\delta$ \citep{2010ApJ...712..405V}. 
The quasars in our sample have a narrow range of $\theta \leq1.1^\circ$. The jets in radio galaxies 0415$+$379 and 0430$+$052 are inclined by 11 and 6 degrees respectively. BL\,Lacertae has $\theta\approx8^\circ$, and is closer to the values for radio galaxies than to the other BL\,Lacs.

\subsection{Acceleration in the jets}
\label{sec:accel}

The core at a given observing frequency is located at different linear distance from the central engine depending on the viewing angle. Therefore, the speeds derived from the core shifts correspond to the different regions along the jet. The de-projected distance of the core from the jet apex at a frequency $\nu$ can be estimated as $R_\nu \approx 8.3\times10^{-8}\Delta r D_A / (\nu\sin \theta)$
(for $k_r=1$, \citealt{1998A&A...330...79L,2011A&A...532A..38S}). 
The estimated de-projected distances of the core at 15 and 8\,GHz are listed in Table~\ref{tab:res}. The difference $R_{8}-R_{15}$ corresponds to the observed core offset in parsecs. The de-projected distance of the core at 15\,GHz varies from $R_\mathrm{15}\ltsim 1$\,pc in radio galaxies to $R_\mathrm{15}>100$\,pc in quasars. 

\begin{figure}
	\includegraphics[width=\columnwidth, trim=0.3cm 1cm 0.3cm 0.3cm]{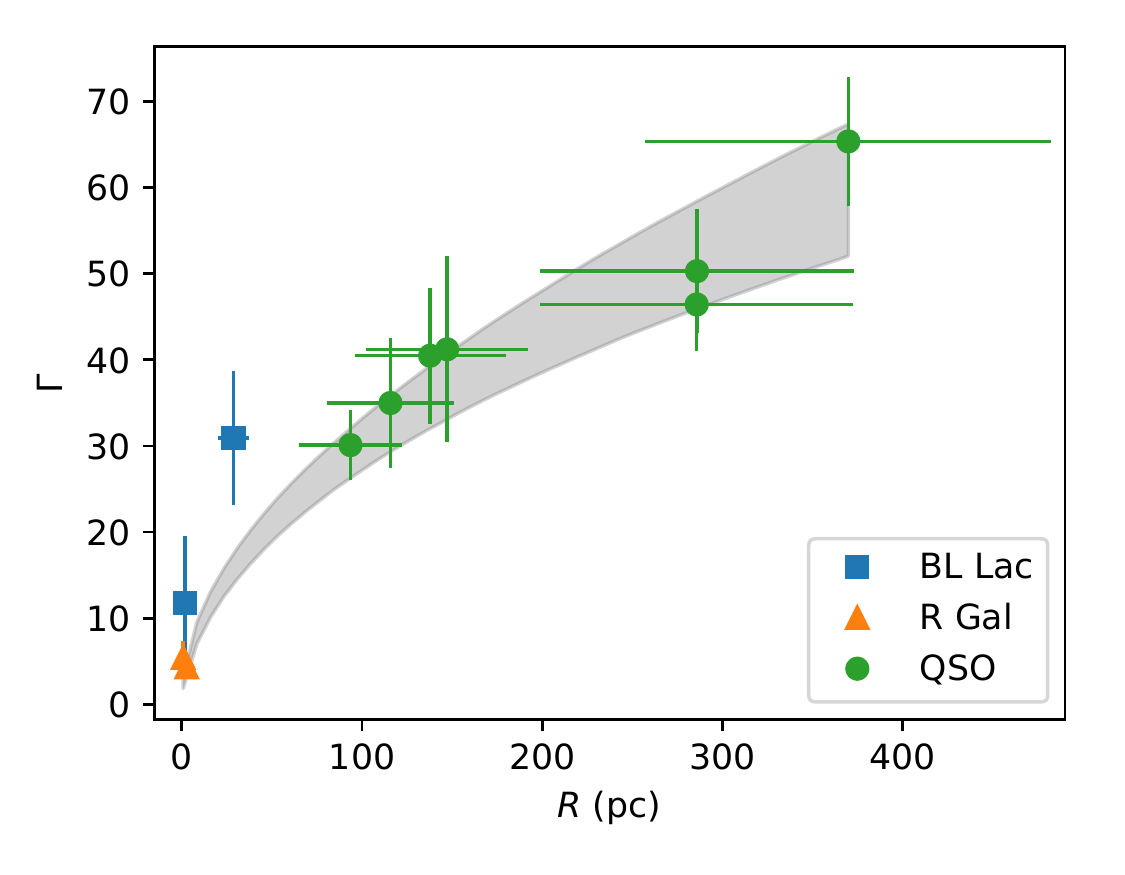}
	\caption{Lorentz factors on various de-projected jet scales. The shaded area shows 95\% of posterior samples obtained with MCMC. The horizontal bars denote $R_{15} - R_{8}$ distance.}
	\label{fig:gamma_rcore}
\end{figure}

In Figure~\ref{fig:gamma_rcore} the Lorentz factor is plotted against the distance probed $R=(R_{15}+R_8)/2$. A simple power-law model with MCMC yields narrow normal posterior distributions of the parameters giving $\Gamma = (2.7\pm0.5)R^{0.52\pm0.03}$ (the errors here refer to a standard deviation of the posterior parameters distributions). The shaded area in Figure~\ref{fig:gamma_rcore} shows the 95\% confidence bounds obtained from the posterior distributions of the Lorentz factors. The BL\,Lacs fall out of the 95\% confidence interval, suggesting that an acceleration in these objects might be faster. We note, that the obtained power law is consistent with the prediction of some MHD models of jets acceleration on those scales~\citep[e.g.,][]{2006MNRAS.367..375B}.

\begin{figure}
	\includegraphics[width=\columnwidth, trim=0.3cm 1cm 0.3cm 0.3cm]{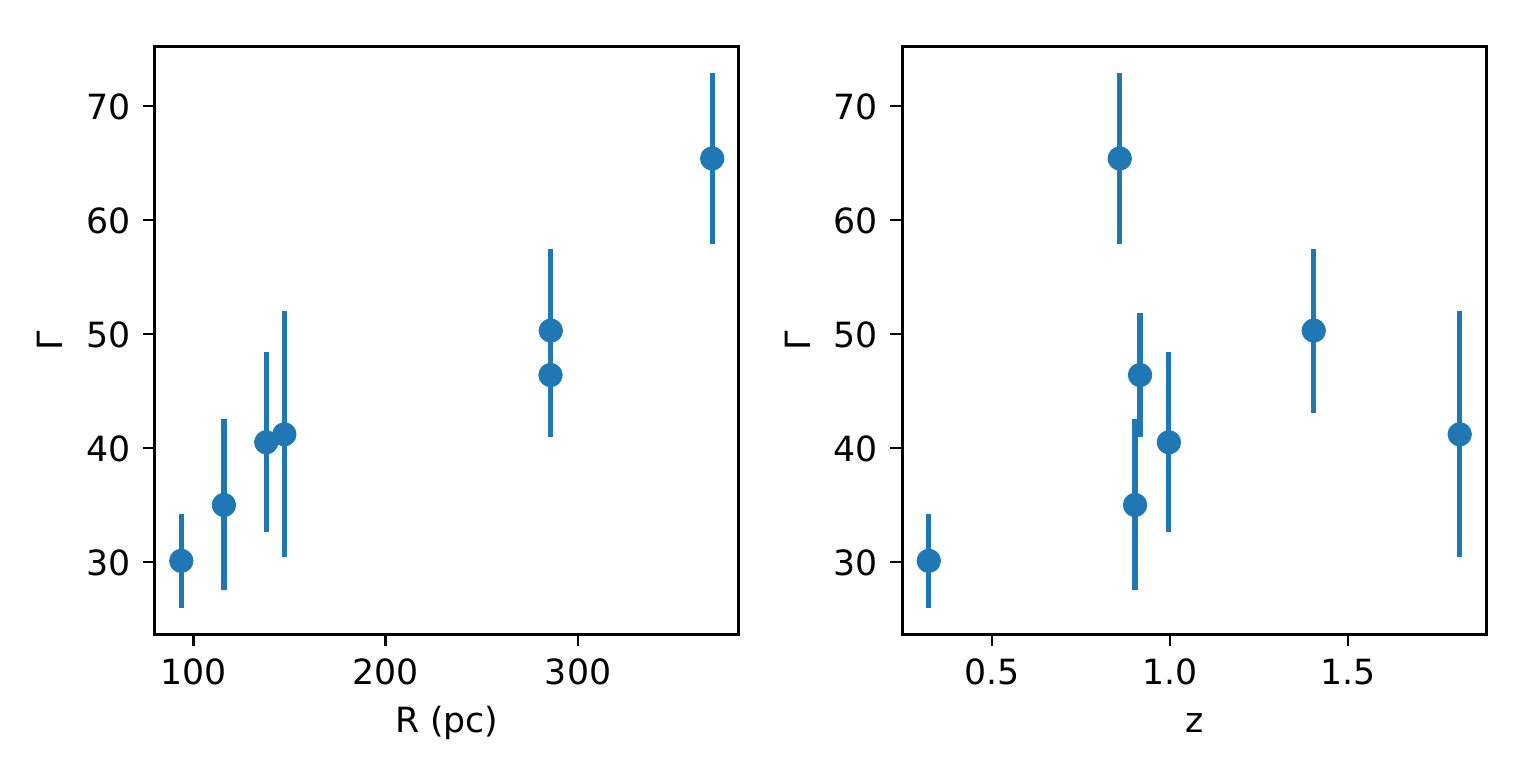}
	\caption{Dependence of the Lorentz factor on the de-projected distance along the jet (\textit{left}) and the redshift (\textit{right}) for the quasars.}
	\label{fig:gamma_Rz}
\end{figure}

We check that this is not a selection effect causing the objects at higher redshifts to have larger Lorentz factors. In Figure~\ref{fig:gamma_Rz} we show the dependence of the Lorentz factor on the de-projected distance along the jet and the redshift for the quasars. As seen in the figure, there is practically no dependence on the redshift, while the Lorentz factor is monotonically increasing with the distance along the jet.   

We come to the conclusion that there is an acceleration of the plasma on de-projected scales 1--500 parsecs. This is in a good agreement with the high resolution observations of the innermost jet scales of nearby AGN~\citep[e.g.,][]{2012ApJ...745L..28A, 2018ApJ...860..141H} as well as with the VLBI kinematical studies~\citep{2009ApJ...706.1253H, 2015ApJ...798..134H, 2017ApJ...846...98J}. Using the core shift and time delays measurements at various frequency pairs one can obtain a detailed acceleration profile of a jet. 
We underline, that the evidence of the acceleration is obtained within the assumption that the jets in our sample are similar, and their main properties are determined by the viewing angle~\citep{1995PASP..107..803U}.

\begin{figure}
	\includegraphics[width=\columnwidth, trim=0.3cm 1cm 0.3cm 0.3cm]{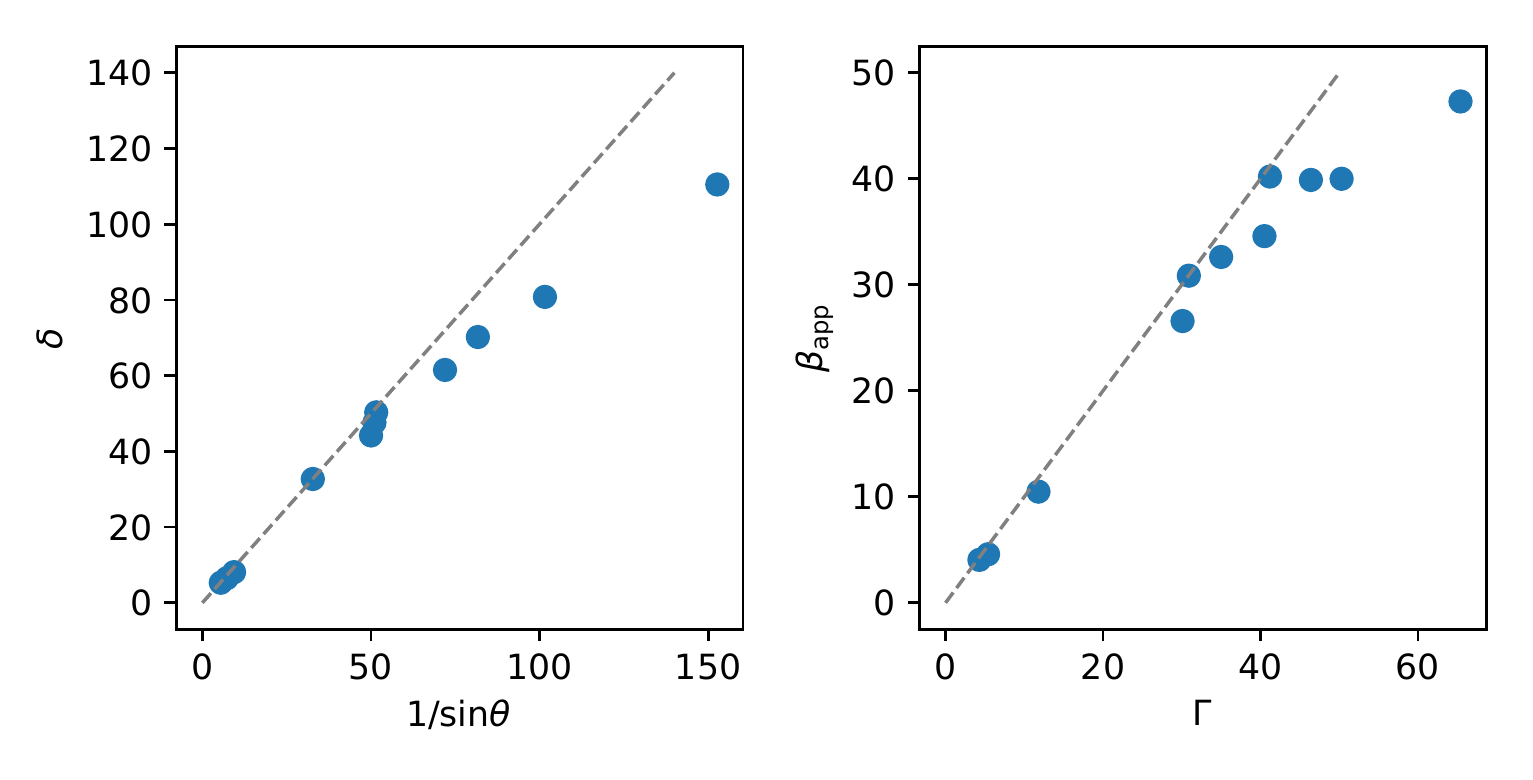}
	\caption{The Doppler factors vs the inverse viewing angles (left) and the apparent speeds vs. the Lorentz factors. The dashed lines show equality.}
	\label{fig:delta_theta}
\end{figure}

In a jet with a given viewing angle the Lorentz factor is increasing downstream resulting in a maximal Doppler factor attained at some region. A source with the core located near that region will demonstrate the strongest flux density variations ($\propto \delta^5$, \citealt{1995PASP..107..803U}). In this case the Doppler factor will be close to maximal possible $\delta_\mathrm{max}\approx1/\sin\theta\approx\Gamma$. And the apparent speed approaches the maximal value of $\beta_\mathrm{app, max} =\beta\Gamma\approx\Gamma$. In Figure~\ref{fig:delta_theta} we show these relations for the estimated parameters. We note that although the viewing angles are estimated using the Doppler factors, the Equations~(\ref{eq:gamma}) do not imply a maximization of any of their components. Therefore these results are self-consistent.

We conclude that the sources in our sample fulfill the condition of maximal Doppler factor in the region of their cores. This is naturally expected since they are among the strongest variable AGN and have been selected based on this criterion. The obtained apparent speeds in the sources are close to their Lorentz factors and are $<50c$. This also explains the result of the correlation between the core shifts and time delays in these objects with roughly the same coefficient $\umu=\Delta r/\Delta t$, indicating that the viewing angle in these sources is near $\theta\sim1/\Gamma$. 
As one can see from Figure~\ref{fig:beta_z}, there are a lot of sources with the highest apparent VLBI speed approaching to our estimates. These jets are expected to obey this relation as well. 
\begin{figure}
	\includegraphics[width=\columnwidth, trim=0 1cm 0 0]{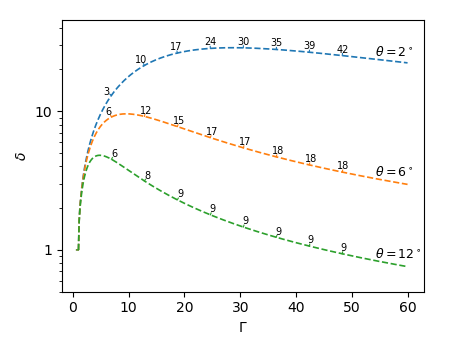}
	\caption{The Doppler factor in an accelerating jet viewed at 2, 6 and 12 degrees. Numbers along the curves show the apparent speed in $c$.}
	\label{fig:bdg}
\end{figure}

Since the core position depends on the observing frequency, the maximal Doppler factor in a jet is attained at some frequency as well (Figure~\ref{fig:bdg}). Hence, in a core-dominated source one would expect the power of flux density variations to change with frequency either monotonically or with a peak, depending on how far the core is from the Doppler factor peak (see Figure~\ref{fig:bdg}). The amplitude of variations of the core is expected to correlate with the apparent speed of the flow, which can be checked on the existing data. 

At least some of the stationary jet features seen by VLBI in many AGN \citep{2018ApJ...853...68P,2017ApJ...846...98J,2016A&A...592A..22H,2015A&A...578A.123R,2013A&A...551A..32F,2012A&A...537A..70S} can also be attributed to the Doppler factor maximization along the jet. They might naturally occur in a source at a frequency where the core is upstream of the position corresponding to a Doppler factor peak. Decreasing the observations frequency will result in that the core will approach a stationary feature and below some frequency the latter will disappear due to the opacity. As seen from Figure~\ref{fig:bdg} these features are expected to be more pronounced in the jets having larger viewing angles. Of couse the other mechanisms responsible for the stationary features might work, e.g. recollimation shocks~\citep{1988ApJ...334..539D}.

\subsection{VLBI components: the brightest one is not the fastest one}
\label{sec:components}

In a jet has a range of Lorentz factors along its extent, the different apparent speeds of the features are naturally explained. Figure~\ref{fig:bdg} illustrates that increasing the Lorentz factor provides a wide distribution of high Doppler factors, corresponding to a vast spread of the apparent speeds. 

The features in radio galaxies have the apparent velocities of $0-10c$, and the highest probability to observe $\beta_{\mathrm{app}}\sim 5c$ provided it is dictated by the Doppler boosting only (see Figure~\ref{fig:bdg}). This is in a good agreement with the observations. 
In a quasar viewed by a small angle, say $2^\circ$, the probabilistic distribution is much wider implying $\beta_{\mathrm{app}} \in [10c; 50c]$ with the highest probability to observe $\beta_{\mathrm{app}}\sim25c$~\citep[see also a discussion by][]{1994ApJ...430..467V}. 
We speculate, that such a picture is expected in an accelerating jet if the features are separated from the plasma flow on various scales (e.g., due to the Kelvin-Helmholtz instability) and then move quasi-ballistically. In this case the measured VLBI apparent speed is determined by the viewing angle and the jet speed at the radius where a component leaves the flow. Moreover, one can see from Figure~\ref{fig:bdg} that the VLBI components with highest Doppler factor (i.e., the brightest) would have apparent speeds lower than the maximal for a jet with a given viewing angle. At the same time, in a nearby source it is possible to detect components moving faster than the plasma speed at the core region (cf. the radio galaxies in our sample, Fig.~\ref{fig:speeds}).
In an accelerating jet the components residing further from the central engine are expected to move faster, which was observed by VLBI at 15 and 43\,GHz~\citep{2009ApJ...706.1253H, 2015ApJ...798..134H, 2017ApJ...846...98J}. 


\section{Summary}
\label{sec:summary}

We use UMRAO 26\,m total flux density monitoring data to measure peak-to-peak time delays of the flares at 15 and 8\,GHz and the variability time scale for 11 sources with high accuracy provided by the Gaussian process regression technique.
The delay shows a strong correlation with the apparent core shift measured by VLBI, providing an evidence for a common nature of these effects. Our results suggest that both the time delay and the apparent core shift correspond to the same de-projected distance in a jet, and are linked via the apparent speed in the core region. This provides a validation of a new method to probe the kinematics of the extragalactic jets based on multi-frequency core shift and total flux density time delay measurements. 

The relation between the apparent core shift and time delay remains similar for different AGN in our sample, which can be explained in terms of maximization of the Doppler factor in the region of the centimeter VLBI core. 
The apparent jet speed in the core region exceeds or equals to the highest velocities obtained from VLBI kinematics analysis in all but one sources. The coefficient between the estimated speed and that of the fastest components traced by VLBI ranges from 0.5 to 20 with median value of 1.4.

We derive Doppler factors, Lorentz factors and viewing angles of the jets, as well as the corresponding de-projected core distance from the jet base. The estimated velocities probe the scales of 0.5--500 parsecs. Our results provide an evidence for jet acceleration on these scales obeying a power law $\Gamma\propto R^{0.52\pm0.03}$.

In an accelerating jet viewed at a given angle a Doppler factor reaches its maximum at some distance from the central engine. A strongly amplified variability is observed from the core when it is located near that distance.  If the core at a given frequency is upstream of the region with maximal Doppler factor, then a stationary component can be observed. A range of the apparent speeds of VLBI components is expected in such a jet. Moreover, the fastest components in a source will be observed further downstream the jet and will not be the brightest in the outflow.

\section*{Acknowledgments}

We thank Peter Voytsik, Eduardo Ros and the anonymous referee for the useful comments which helped to improve the manuscript.
This research was supported by Russian Science Foundation (project 16-12-10481).
UMRAO was supported in part by a series of grants from the NSF, most recently AST-0607523, and by NASA Fermi Guest Investigator grants NNX09AU16G, NNX10AP16G, NNX11AO13G, and NNX13AP18G. 
This research has made use of data from the MOJAVE database that is maintained by the MOJAVE team~\citep{2018ApJS..234...12L}.

\bibliographystyle{mnras}
\bibliography{refs}

\label{lastpage}
\end{document}